# Sub-Millisecond Visible-Light Gating of a Zinc Oxide Nanowire

Justin Derickson[1,+], Benjamin Kerr Barnes[2,3+], and Kausik S Das[2,*,+]

[1]Department of Engineering, University of Maryland Eastern Shore, 1, Backbone Road, Princess Anne, MD 21853 USA
[2]Department of Natural Sciences, University of Maryland Eastern Shore, 1, Backbone Road, Princess Anne, MD
21853 USA
[3]Department of Chemistry and Biochemistry, University of Maryland College Park, MD 20742 USA

*Correspondence and requests for materials should be addressed to K.D (kdas@umes.edu)

+these authors contributed equally to this work

ABSTRACT

Semiconductor nanowires are the building blocks of many nanoscale electrical and neuromorphic circuits. Here, we demonstrate a simple arrangement wherein an ethanol-adsorbed ZnO single nanowire, deposited between gold electrodes using dielectrophoresis exhibits significant change in resistance when activated by visible light. Here we have observed that the transition timescale between two stable ohmic states, one in the dark and the other in the illuminated regime in a single nanowire can occur in the submillisecond order, which is 7 orders of magnitude lower than our previously reported switching timescale in bulk ZnO thin films. A mathematical model of the light activated resistance switching mechanism is proposed based on the adsorption-desorption kinetics of oxygen molecules at the surface of the nanowires.

TEXT

Zinc oxide is a promising material for alternative memory devices due to its strong photoresponse to UV illumination[1,2]. Previous works have shown resistance switching in ZnO with visible light activation by decorating the surface with quantum dots[3,4] or plasmonic metal nanoparticles[5]. Recently we reported the observation of visible light induced memristive photoswitching in bulk ZnO films which we qualitatively attributed to light-induced desorption of oxygen molecules from the ZnO surface[6]. However, a proper mathematical model was not developed in that paper. We found that soaking the ZnO material in ethanol during the fabrication process enabled this photoswitching effect under visible light illumination. Other works have also shown sub-bandgap illumination to be capable of inducing a photocurrent in native ZnO which was attributed to hole traps[7]. However, ZnO films used in previous studies, including ours have exhibited a slow photorepsonse, requiring minutes to reach the saturation value, which would not be suitable for memory device applications where rapid switching is required. It is desirable to maintain the visible light induced switching while decreasing the response time, which would enable further progress in ZnO in memory devices[8,9,10,11,12,13,14,15], gas sensing[16], and memristive computing[17,18,19,20].

Since the visible light induced photoswitching observed in our work is hypothesized to be due to the kinetics of oxygen adsorption and desorption, we anticipate that the switching time is proportional to the surface area of the ZnO channel, so the use of a single nanowire will yield a decreased switching time. Furthermore, by restricting electronic conduction to a smaller ZnO channel by the use of a nanowire, we can anticipate an amplified signal compared to the bulk film devices we explored earlier. In this work, we demonstrate that by using a single ZnO nanowire deposited from ethanol, we can achieve sub-millisecond photoswitching under visible light illumination. In our previous work we have showed explicit dependence of wavelength on ZnO Furthermore, we present a mathematical model which successfully describes the photoswitching rate in terms of oxygen adsorption and desorption.

Single nanowire bridges between two electrodes of varying gaps were created by dielectrophoresis of ZnO ethanol mixture on a gold electrode array[21]. 16 mg of ZnO nanowires were dispersed in 1 mL of ethanol with several minutes of bath sonication, and diluted with 1 mL of water. Controlled positioning of ZnO nanowires between the electrodes was achieved by dispensing a 10 µL drop of nanowire solution between the electrodes on the wafer and applying an alternating voltage of 20 V peak to peak ($V_{pp}$), centered at 0 V at 10 MHz frequency for a few minutes. After the solution was dispensed and the alternating electric field was applied between the electrodes we used an optical microscope to watch the individual nanowires oscillating in the field and eventually making bridges between the electrodes (Figure 1). Once the nanowires were pinned between the electrodes, remaining solution was blown off by compressed air and then the function generator was turned off. Some devices were observed with multiple nanowires, but these were not included in the study. No agglomerated nanowires were observed bridging the electrodes. This repeatable process enabled us to make several ZnO nanowire bridges between electrodes of varying gap widths down to 2 µm for electrical characterization.

Electrical characterization was done in two ways (Figure 2), firstly via current limited method, where current was kept constant at 10 mA using a Kiethly 2450 source meter. Connections with the electrodes were established using micromanipulator tips (7F Tungsten Cat Whisker Probe Tip, The Micromanipulator Co.) and the potential drop across the nanowire bridge was measured using a standard oscilloscope. Secondly, in voltage limited situation both the input voltage and the output current were measured by the source meter.

When illuminated with visible light from a solar simulator (Abet Technologies, 1 Sun output), the resistance switching times between the high resistive state (HRS) and a low resistive state (LRS) in the range between 700 µs to 8 ms are observed for ZnO nanowire bridges of varying length and diameters. Wratten filters of different colors show that qualitatively resistance switching phenomena in a single ZnO nanowire doesn't depend on the wavelengths of the visible light, similar to the observation in bulk ZnO thin films[6]. We used both current limited and voltage limited schemes to investigate and record the light activated resistance switching phenomenon. Figure 3 shows a current limited resistance switching when the current is kept constant at 10 µA through the nanowire. Change in the potential drop across the nanowire is measured using an oscilloscope while a light chopper creates periodic dark and illuminated conditions on the nanowire. We have also noticed that there is a slight difference in growth (~ 2 ms) and decay (~ 400 µs) time constants in the same nanowire (Figure 3 a, b), pointing to a difference in time scales between the absorption and desorption kinetics.

We have also developed a mathematical model based on the hypothesis that in absence of light the ZnO nanowire, being a naturally n-type semiconductor, adsorbs oxygen molecules at the active sites on its surface[22], which, upon exposure to light, get displaced and desorbed[23,1]. These adsorbed oxygen molecules trap free electrons from the conduction band of the nanowire[24,25], thereby, reducing the number density of the effective charge carriers (Figure 4 a). As light shines on the nanowire the weakly adsorbed oxygen molecules are desorbed releasing the trapped electrons back to the conduction band (Figure 4 b).

This means that the Fermi level shifts towards the conduction band and the conductivity of the nanowire increases due to the increased density of free charge carriers. When the light is turned off, atmospheric oxygen starts to re-adsorb on the surface of the nanowire again and electrons start to get trapped by these chemisorbed oxygen. Fermi level starts to shift towards the valence band at this point.

When a ZnO nanowire is exposed to light for a prolonged amount of time its resistance decreases to a steady state value. At this time the charge density of electrons in the nanowire is maximum and the number of surface trapped oxygen molecules are minimum. The steady state charge density of electrons in a single ZnO nanowire in the presence of light for $t \to \infty$ is written as $n_e^\infty$. Obviously, the steady state current under light illumination is $I_{il} = n_e^\infty A v_d e$, where $A$ is the cross sectional area of the nano-wire, $e$ is the electron charge and $v_d$ is the drift velocity of the untrapped charge carriers. At this asymptotic limit the least number of electrons are trapped by the surface adsorbed oxygen molecules. As the light is turned off oxygen molecules start to re-adsorb on the nanowire surface and the number of electrons trapped by the surface adsorbed oxygen $n_{tr}(t)$ starts to increase, in turn decreasing the number density of the charge carriers contributing to the current flow.

We start with a ZnO nano-wire kept in prolonged dark conditions and then light is turned on at $t = t_{on}$. Let us suppose that in the steady dark state the number density of saturated O₂ molecules adsorbed on the surface of the nanowire is $n_{O_2}^d$, which is practically a constant. As the nanowire is exposed to light, the surface adsorbed O₂ molecules start to desorb. We hypothesize that, similar to the law of mass action, the rate of desorption is proportional to the number density of O₂ molecules adsorbed on the surface of the nanowire, i.e.,

$$\frac{d}{dt} n_{O_2}^S(t) \propto n_{O_2}^S(t) \qquad (1)$$

Leading to

$$\frac{d}{dt}n^S_{O_2}(t) = -k_d n^S_{O_2}(t) \qquad (2)$$

where $K_d$ is the desorption constant and $n^S_{O_2}$ is the number density of surface adsorbed oxygen molecules at a given time $t$. $K_d$ in general should depend on temperature, light intensity, radius of the nanowires etc.

We know that the adsorbed $O_2$ molecules are responsible for trapping the negative charge carriers from the conduction band[6]. A plausible hypothesis is that the number density of surface trapped charge carriers is directly proportional to the number density of surface adsorbed oxygen molecules, and can be written as

$$n^{tr}_e(t) = \alpha_d n^S_{O_2}(t) \qquad (3)$$

where $\alpha_d$ is a coupling constant which depends on the surface properties of the nanowire and $n^{tr}_e(t)$ is the number density of surface trapped electrons at an instant t after the light is turned on at time $t_{on}$. Suppose after sufficiently long time, i.e., at $t \to \infty$, the desorption of $O_2$ molecules reach saturation and the number density of free electrons reach the steady state value $n^\infty_e$. So, the saturated steady current under illumination can be written as $I_{il} = n^\infty_e A v_d e$ where $A$ is the cross sectional area of the nano-wire, $e$ the electron charge and $v_d$ is the drift velocity of the untrapped charge carriers. After the light is turned on the instantaneous current under illumination is thus nano-wire, $e$ the electron charge and $v_d$ is the drift velocity of the untrapped charge carriers. After the light is turned on the instantaneous current under illumination is thus

$$I(t) = n_e(t) A v_d e \qquad (4)$$

where $n_e(t) = n^\infty_e - n^{tr}_e(t)$ is the number density of free electron carriers responsible for the flow of current, $n^\infty_e$ being the number density of free electrons at the steady illumination state as at $t \to \infty$ after the light is turned on, the desorption of $O_2$ molecules reach its saturation. Hence, the current in the nanowire can be written as

$$I(t) = \left(n_e^\infty - n_e^{tr}(t)\right) A v_d e \tag{5}$$

And thus,

$$\frac{dI}{dt} = -\frac{dn_e^{tr}}{dt} A v_d e \tag{6}$$

Using Eq. (2) and Eq. (3) we get

$$\frac{dI}{dt} = K_a n_e^{tr} A v_d e \tag{7}$$

Using Eq. (6) and the expression $I_{il} = n_e^\infty A v_d e$ we get

$$\frac{dI}{dt} = K_a \left(I_{il} - I(t)\right) \tag{8}$$

After integrating both sides of Eq. 8 with the initial conditions that at $t = t_{on}$, $I = I_d$ the solution can be written as

$$I(t) = I_{il} - (I_{il} - I_d) \exp\left(-K_a(t - t_{on})\right) \tag{9}$$

where the current starts from $I_d$ when the light is turned on and reaches the asymptotic value $I_{il}$ as $t \to \infty$. A comparison of the theoretical result (Eq. 9) and the experimental data is shown in Fig. 5. Using the same technique we can also show that the decay of current follows the equation below:

$$I = I_d + (I_{il} - I_d) \exp\left(-K_{ad}(t - t_{off})\right) \tag{10}$$

where $I_{il}$ is the saturation current under illuminated condition, $I_d$ is the dark saturation current, $t_{off}$ is the time when the light is turned off and $K_{ad}$ is the adsorption coefficient.

Plotting this model vs time yields an excellent fit to the experimental data (Figure 5).

In conclusion, we have shown that an ethanol-adsorbed single ZnO nanowire shows resistance switching, which in principle can be used as a building block for neuromorphic circuits. The time scale of the resistance switching in this case is in the order of milliseconds to a few hundred microseconds, which is orders of magnitude less than the resistance switching time scale of a ZnO thin film as previously reported[6]. We have also developed a mathematical model based on the hypothesis that surface adsorbed $O_2$ molecules are primarily responsible for the resistance switching in ZnO. We have shown that the adsorption-desorption kinetics of $O_2$ molecules and the resulting trapping of electrons from the conduction band electrons can determine the free charge density of the nanowire in presence/absence of light and the dynamics of current in it in time. The theoretical model shows good agreement with the experimental observations.

AUTHOR CONTRIBUTIONS

K.D. conceived the idea, the experiment strategies, the implementation plan and supervised the project. He also designed the study, analysed and interpreted data, and wrote this manuscript. B.B prepared ZnO nanowire bridges using dielectrophoresis technique. JD set up the experiment and improvised many a times to optimize the experimental set up. JD also collected data and jointly analyzed them with KD. All authors were involved in writing and reviewing the manuscript.

ACKNOWLEDGEMENT

This work was partially supported by the National Science Foundation (Award # 1719425), the Department of Education (MSEIP Award # P120A70068) and Maryland Technology Enterprise

Institute through MIPS grant. KD would like to thank Dr. Jim Marty of Minnesota Nano-Science Center and NanoLink and Prof. J.K. Bhattacharjee for many helpful discussions and suggestions.

DATA AVAILABILITY

The data that support the findings of this study are available from the corresponding author upon reasonable request.

FIGURE CAPTIONS

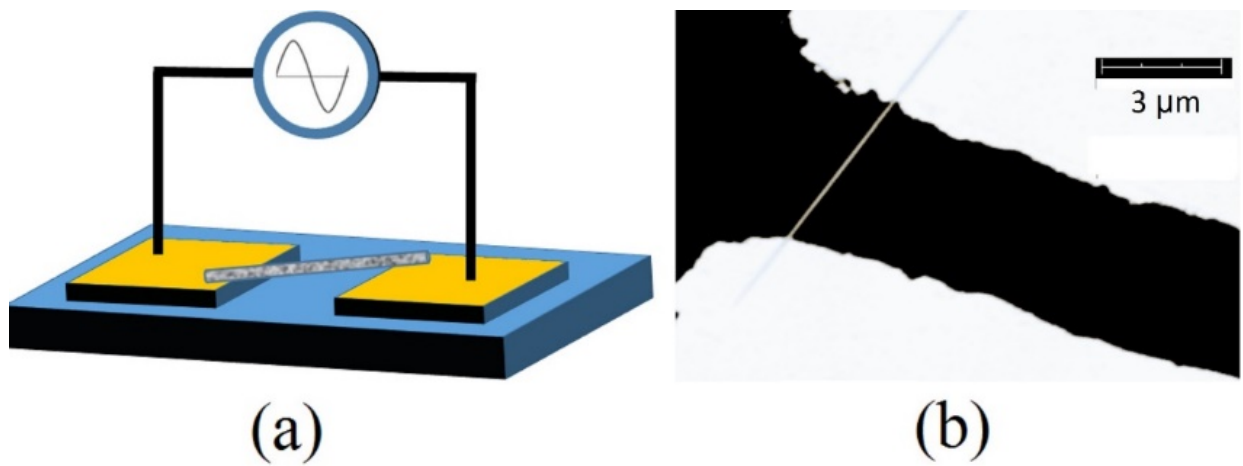

Figure 1. ZnO nanowire bridges between the electrodes are created using dielectrophoresis technique. A single ZnO nano-wire bridge is formed between the electrodes. Several single bridges are formed between the electrodes of gap width raging between 2µm and 10µm. (a) An alternating voltage of 20Vpp is applied between the gold electrodes and a drop of dilute suspension of ZnO in ethanol is placed between the electrodes. ZnO nanowire bridges are formed by this dielectrophoresis technique. (b) SEM image of a single nanowire bridge formed between the electrodes.

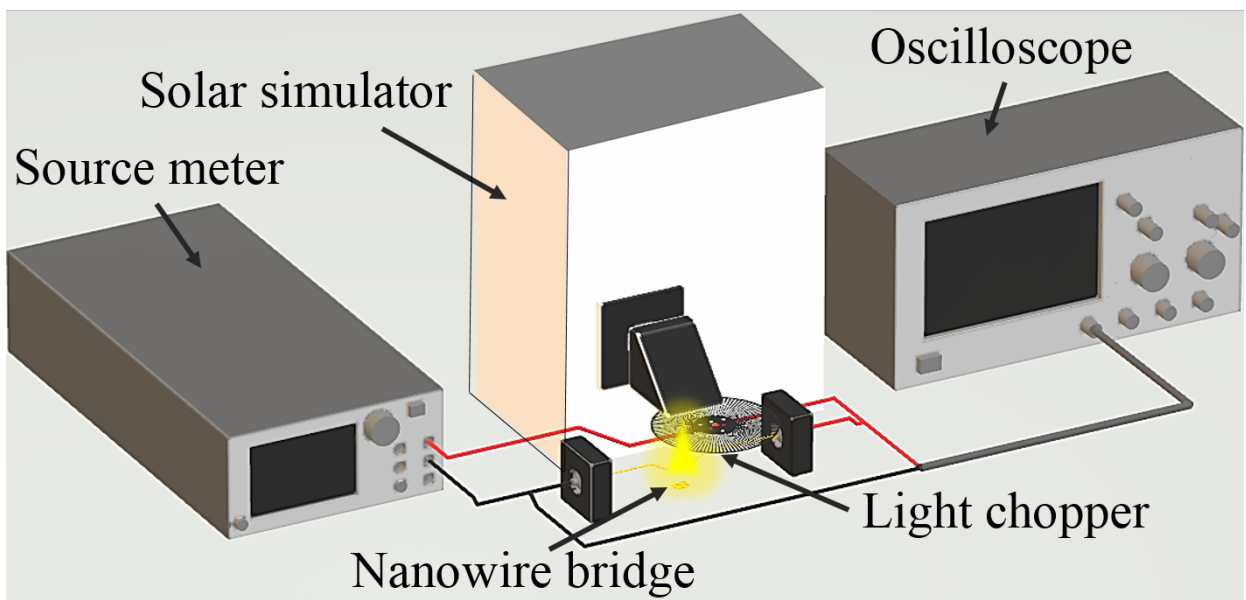

Figure 2. Experimental set up for the measurement of resistance switching in a single ZnO nanowire. The light chopper in between the solar simulator and the nanowire bridge illuminates the nanowire periodically, depending on the light chopper rotation speed. Each blade of the light chopper wheel creates a shadow or dark region on the nanowire bridge, while the gap between two consecutive blades create an illuminated region. The rotation speed of the light chopper can be adjusted by a controller. In the current limited system the input is controlled by a source meter and the optoelectrical response of the nanowire is measured by an oscilloscope wired to the device by electrical microprobes.

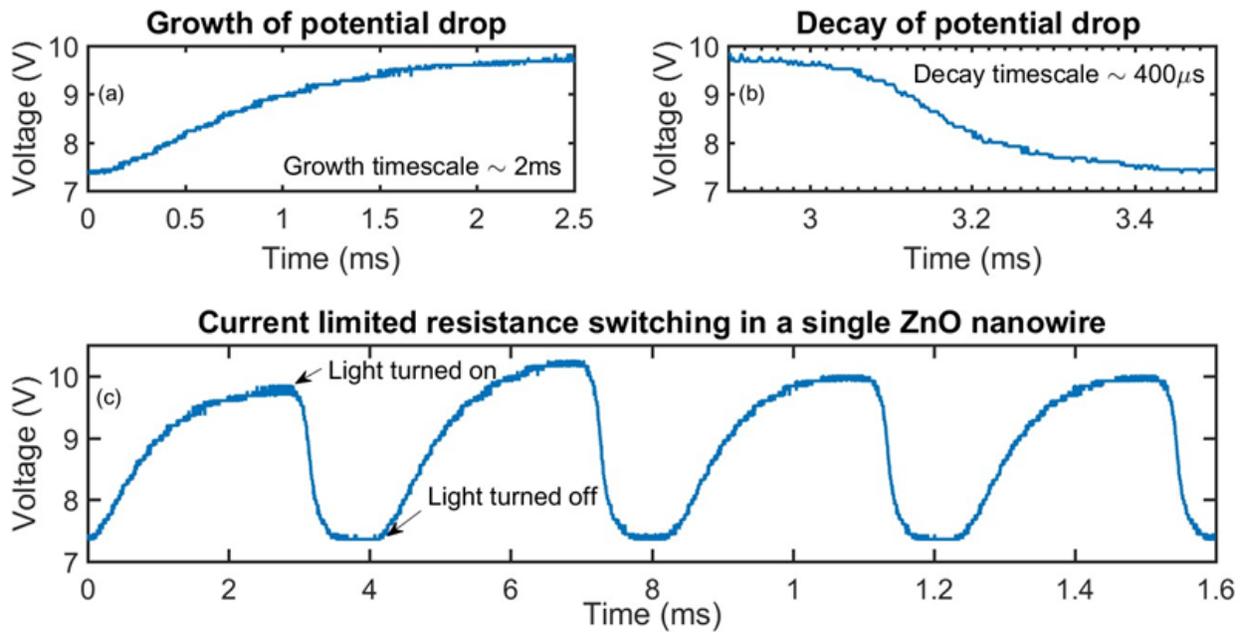

Figure 3. Potential drop across a 2µm long nano-wire bridge switches periodically as the light is turned on and off periodically. Current is kept constant at 10 mA. (a) shows the growth of the potential drop across the nanowire when light is turned off from a steady state illuminated state. The growth represents an increase in resistance of the nanowire and a transition from the HRS to LRS. (b) shows a decay in potential drop in a current limited situation, indicating a transition from HRS to LRS.

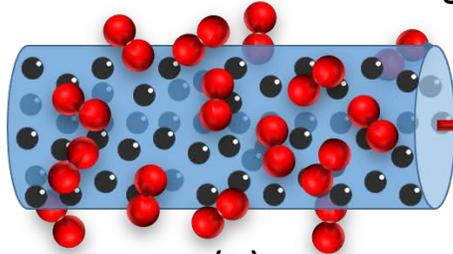

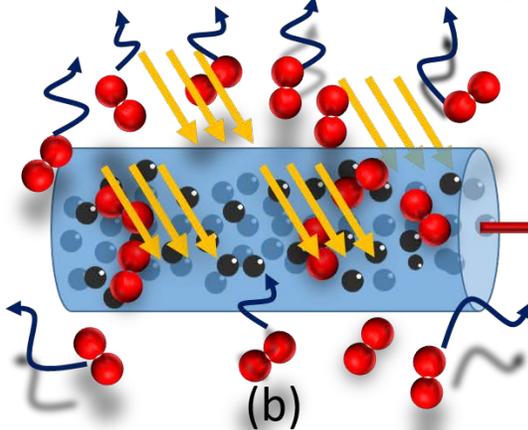

Figure 4. Resistance changing mechanism based on charge carrier trapping by surface adsorbed Oxygen molecules. (a) In prolonged dark condition surface adsorbed Oxygen molecules trap free electrons thereby reducing the density of free charge careers. (b) Exposure to light causes gradual desorption of the chemisorbed Oxygen molecules, thereby releasing the trapped electrons back to the conduction band of the nanowire. Resistance decreases gradually and goes to the asymptotic limit of LRS as $t \to \infty$.

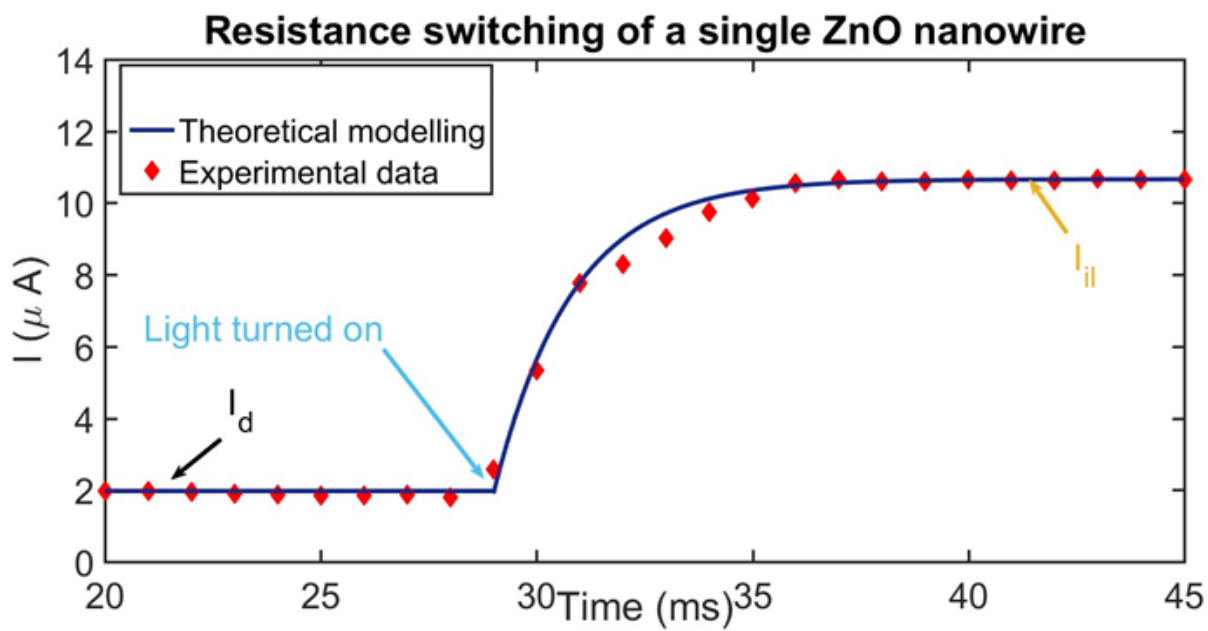

Figure 5. Experimental observation of resistance switching of a single ZnO nanowire compared with the theoretical results. A constant potential difference of 20 V was applied across the nanowire bridge. Resistance starts to drop and consequently current starts to increase as light is turned on.